# UNIPOLAR MAGNETIC FIELD PULSES AS TRANSIENT SIGNALS PRIOR TO THE 2009 AQUILA EARTHQUAKE SHOCK

P. Nenovski

UCSRT, Sofia University, Sofia, Bulgaria

**Abstract**. Unipolar pre-seismic magnetic field pulses have been observed first by Bleier et al. (2009) and Villante et al. (2010) and Nenovski et al. (2013). In the present study a detailed analysis of the pulses is conducted looking for signatures of transient signals similar to that recorded at the 2009 Aquila earthquake main shock (Nenovski, 2015). Various magnetic field data around the Aquila earthquake provide an instrumental basis for such an analysis. In addition to the fluxgate magnetometer data (already examined in previous studies), overhauser magnetometer data are involved. The result is a detection and discrimination of transient signals of diffusive form that appear prior to the earthquake shock.

*Keywords:* transient signal, diffusive form, pre-seismic and co-seismic, signal polarity, pulse width,

**Introduction**

Attempts to examine electric and magnetic field effects caused by irreversible mechanical. processes as cracks, fractures and stick-slip events at the time of earthquakes have been conducted for long time. A lot of mechanisms of generation and release of electric charges in rocks and hence are considered, then electric currents are expected as a underlined mechanism of electric and magnetic fields generated prior and/or during earthquakes. For example, step-like changes in stress and strain in rocks should result in an enormous electrical. signal. because of its short duration at the time of indentation fracture (Enomoto and Hashimoto, 1990, 1992, Parrot, 1995). On the other hand, unipolar pulses recorded by search-coil (induction) magnetometers have been recorded preceding earthquakes (Bleier et al., 2009; Bortnik et al., 2010; Villante et al., 2010, Nenovski et al., 2013). During the weeks leading up to the M = 5.4 Al.um Rock earthquake of 30 October 2007, a magnetometer located about 2 km from the epicenter recorded unusual non-al.ternating magnetic pulses, reaching amplitudes up to 30 nT (Bortnik et al.., 2010; Bleier et al.., 2009). The incidence of these pulses increased as the day of the earthquake approached. Note that after the earthquake, the pulse count immediately falls back to normal. levels. These local.ized unipolar pulses were much stronger and lasted longer, and could not be seen at locations far from the EQ epicenter (Bleier et al.., 2009). Villante et al. (2010) have observed an event of unusual. clusters of unipolar pulses of amplitude ~ 1 nT and width up to several. seconds on March 18, 2009, 19 days before the Mw 6.1 2009 Aquila earthquake. Their mechanism however has not been considered. Earlier, on 14 and 16 February 2009, unipolar pulse activity also has been identified, it lasted up to several hours under night-time conditions (Nenovski et al., 2013). It was concluded that the nature of the pulse signatures prior to the L'Aquila earthquake is unknown and its possible relationship to the L'Aquila earthquake (it occurred 55 days later) is not easy to be proved.

Until recently, field observations of electric and magnetic signals at the time of earthquakes however remains still rare. The only exception were the piezomagnetic changes in the total. magnetic field related to stress (Stacey, 1964; Johnston et al., 1994; Johnston, 2002). Recently, single signals of unipolar sense that started just after the rupture of the fault have been detected first by electric field measurements (Fujinawa et al., 2011) and very recently by magnetic field measurements (Nenovski, 2015). To be exact, co-electromagnetic signals just before the seismic wave arrival (about 10 seconds and ~4 seconds after the EQ shock time) have been recorded first by Belov et al. (1974). Belov et al. (1974) however have measured only ULF field intensity filtered in specified frequency diapason and have not studied their characteristics. Authentic electric field pulses have been detected in Japan by borehole antenna in the 0.01-0.7 Hz channel (Fujinawa et al., 2011). The observed pulse width seemingly exceeds the seismic wave periods, but lies within a minute and less (Fujinawa et al., 2011). The transient magnetic field signal. of 0.8 nT amplitude of unipolar type and pulse width that exceeds 1 minute and overal.l duration of up to 4(5) minutes was observed just after the earthquake shock. The latter was registered at ~10 km distance from the 2009 L'Aquila earthquake hypocenter (Nenovski, 2015).

In contrast to the co-seismic magnetic field signal. recorded at the Aquila earthquake, the magnetic field unipolar pulses preceding the earthquake shock were of much shorter duration and usually does not exceed several. seconds (Bleier et al., 2009; Villante et al., 2010, Nenovski et al., 2013; Scoville et al., 2015). Another specific feature of the observed magnetic field signals is their amplitude shape: the amplitude increase in very short time, while after the amplitude peak the amplitude decrease last for much more time (Bleier, 2009; Scoville et al., 2015; Nenovski, 2015).

The pulse activity prior to earthquakes observed by Bleier et al. (2009) and Scoville et al. (2015) and the co-seismic pulse at the earthquake shock (Nenovski, 2015) put the following question: is there a common underlying mechanism of the pulse activity prior the earthquake and the transient co-seismic signal? Farther, if the mechanism is what is the relationship with the earthquake processes

Starting point of the present study is a detailed analysis of the unipolar pre-seismic pulses and their structure looking for signatures of transient signals. Various magnetic field data around the Aquila earthquake provide such a study. In addition to the fluxgate magnetometer data (already examined in the previous studies), the overhauser magnetometer data are involved in the analysis and transient unipolar signals prior to the earthquake shock are discovered.

## Data and Results

For the present analysis data from fluxgate and overhauser magnetometers are used (http://roma2.rm.ingv.it/it/risorse/banche_dati/39/osservazioni_relative_al._sisma_del_6-4-2009_a_laquila). The needed instrumentarium is well described by Palangio (2009) and also in Villante et al., 2010; Nenovski, 2015. The magnetic field pulse activity as observed by the two fluxgate magnetometers at L'Aquila, are mostly of dichotomous type or spikes of variable width ranging between 1 and 16 seconds. Figure 1 illustrates the pulse activity recorded by fluxgate magnetometers on March 18, 2009 (Villante et al., 2010). The pulse activity appears as clusters of several. pulses of different width and amplitude. The amplitude is different in components and reaches 1 nT.

Figure 2 includes overhauser magnetometer data. One sees a good correspondence between the some pulses and transient signals marked by ellipses. At 14:32 UT there are two consecutive transient signals, the first one is of 1.2 nT peak amplitude, the second one is of some less amplitude. Their width varies: about 4 and 1 seconds, respectively. Another sequence of transient signals can be assumed by hints of pulses (denoted by the second ellipse). They look like teeth.

The next figure (Figure 3) yields the observed transient signal. in an exaggerated form. The stepped form visible around the base line (F = 46378.2 nT) at that time is an indication that the signal. amplitude is close to the instrument threshold sensitivity. The signals give the impression that the signals if of transient nature are very similar to that examined by Nenovski (2015).

Next step is a detailed study of the magnetic field derivative: dB/dt. The magnetic derivative of the signal is sketched in Figure 4 (the black dash line). The reason is twofold. The first one is linked with the magnetic field measurements by itself. The search-coil (induction) magnetometer records essentially the magnetic field derivative; the second one is related to a better identification of the diffusion process as an underlying mechanism of the transient signal already claimed by Nenovski (2015). As it can be seen form Figure 4, the experimental. magnetic field derivative represents a negative pulse followed by a positive bulge. This means that such an amplitude profile should be a characteristic and important signature of transient processes of diffusive from.

The next step of analysis is to see whether the transient signals resemble a diffusion process. The fitting procedure is done by a semi-quantitative diffusion model and the results are plotted in Figures 4 and 5. The model simulates the magnetic field and its derivative produced by a source of electric current channel of infinite length. The assumed current channel is of arbitrary thickness while the current pulse is initiated at some initial moment and disappears then (a delta function source). A good coincidence for both the magnetic field and its derivative may be obtained for certain diffusion scale. Note that the experimental data indicates an offset of the base line (see Figures 3 and 5). If this offset was ignored (see the dashed line, Figure 5) the fit becomes even better.

It is worth noting that Scoville et al. (2015) have elaborated a model to explain the pre-earthquake unipolar pulse characteristics – its unipolarity, form and duration. In their model the assumed mechanism is p hole generation process, under which charges are pulled and attracted depending on the stress changes. In fact, Scoville et al. (2015) have assumed a process of initiation of p holes under stress (known as Freund mechanism (2002)). The magnetic field signals then appear as a consequence of diffusion of p holes repelled through the boundary between the stressed and unstressed volumes, i.e. the pulse shape has to reflect some diffusion processes. The attenuation of the magnetic field as it passes through the Earth however is not considered, nor are the effects associated with the surface of the Earth (Scoville et al., 2015). Interestingly, Scoville et al. (2015) have intended to interpret the unipolar pulses recorded by search-coil data which yield magnetic field derivative variations. Therefore if the assumed process of generation of the pulses is of diffusion type, the expected pulse shape of the magnetic field derivative would possess an unipolar pulse followed by a bulge of opposite sign. As a result, the experimental. pulses should not be purely unipolar ones, as Scoville et al.. (2015) referred to (see their Figure 1).

**Discussion**

The assumption of electric current source as an underlying mechanism of the observed magnetic field pulses is straightforward (Bortnik et al., 2010; Nenovski et al., 2013; Scoville et al., 2015). It is well known that under earth conductivity conditions, e.g. $\sigma = 10^{-1} \div 10^{-3}$ S/m, the electric current, once excited, would attenuate quickly (miliseconds and less). The observed pulse width of longer periods, say of several. seconds seems to be unexplainable in terms of electric current source except the case when the current source persists al.l the time. An alternative explanation is the magnetic field diffusion effect when the signal. is generated far away from the measuring point. The electric current source was suggested by Nenovski et al. (2013) in order to explain the magnetic pulse polarization characteristics. The observed pulse activities (both on 16 February and on 18 March, 2009) possess similar polarization properties. Their magnetic field orientation in horizontal. plane occurred in NE-SW direction at about 36 degrees from the North. This orientation proves nearly perpendicular to the fault strike that is oriented in NW-SE direction. This finding suggests that the pulse activity has a common source − electric currents generated within the crust being of different amplitudes and duration (continuous series (14 and 16 February) or in clustered form (18 March)). Should the pulses have an earth current genesis, then the observed polarization angle would be indicative for the electric current orientation. Having in mind that the EQ epicenter was practically in the South direction from the L'Aquila magnetic observatory, one may conclude that the current orientation have an angle of less than 20 degrees to the main Apennine fault (NW-SE direction) (Nenovski et al., 2013). The analysis of the polarization of the transient signals yields an angle of −36 degrees to North). This implies that the assumed electric current would possess an orientation closer to North than the fault strike but falls within the strike orientation within a cone of 10 degrees. Thus the geometry of the underlying mechanism whatever it is should be aligned close to the fault strike. Complex analyses of fault structures, 3-D conductivity distribution, different orientation, depth, and extent of the current and other factors (as boundaries) need to be taken precisely into account in order to identify possible characteristics of the current source (Nenovski et al., 2013).

Our analysis demonstrates first, the following common characteristics between the pre- and co-seismic signals: a unipolarity and pulse shape (envelope) of diffusive form. Second, the difference between the two types of pulses (pre-seismic and co-seismic ones) consists in their width: up to several. seconds for the pre-seismic and over 1 minute for the co-seismic one; Third, we succeeded to detect and discriminate an appearance of transient signals of diffusive form prior to and at the earthquake shock.
Generally speaking, observational evidences of unipolar pulses associated with pre earthquake processes are not widespread as they should be. The basic reason is: for most pulses the pulse widths falls within one second and less (Bortnik et al., 2010; Bleier et al., 2009). Therefore the pulses activity (when lies above 1 Hz) cannot be recorded by the conventional global magnetometer networks (with 1 Hz sampling frequency). In addition, the fluxgate magnetometers data alone seems to fail in detecting safely the diffusive properties of the unipolar pulses (Nenovski, 2015). This explains why these pulses have not been detected and discriminated till now.

**Conclusion**

In this study we examined the pre-seismic pulse activity observed before the 2009 Aquila earthquake using both the fluxgate and overhauser magnetometer data. Transient signals of diffusive form were revealed probably indicating a common underlying mechanism, namely an electric current source probably oriented along the fault strike. Note that the magnetic field signals before the Aquila earthquake is nearly perpendicular to the strike whereas the transient signal recorded at the earthquake shock is nearly parallel to the same strike.

Another important issue of this study is the specific temporal form of the observed transient signals. The temporal patterns of the magnetic field and its derivative would be of practical. interest in seeking for pre- and co-seismic signals exploiting flux-gate and search-coil magnetometer data sets, respectively. This intension would be achieved only if the conventional. sampling frequency (of 1 Hz) will be enhanced drastically. Their specific amplitude form, pointed out in our analysis would facilitate the searching procedure and enable their identification.

FIGURES

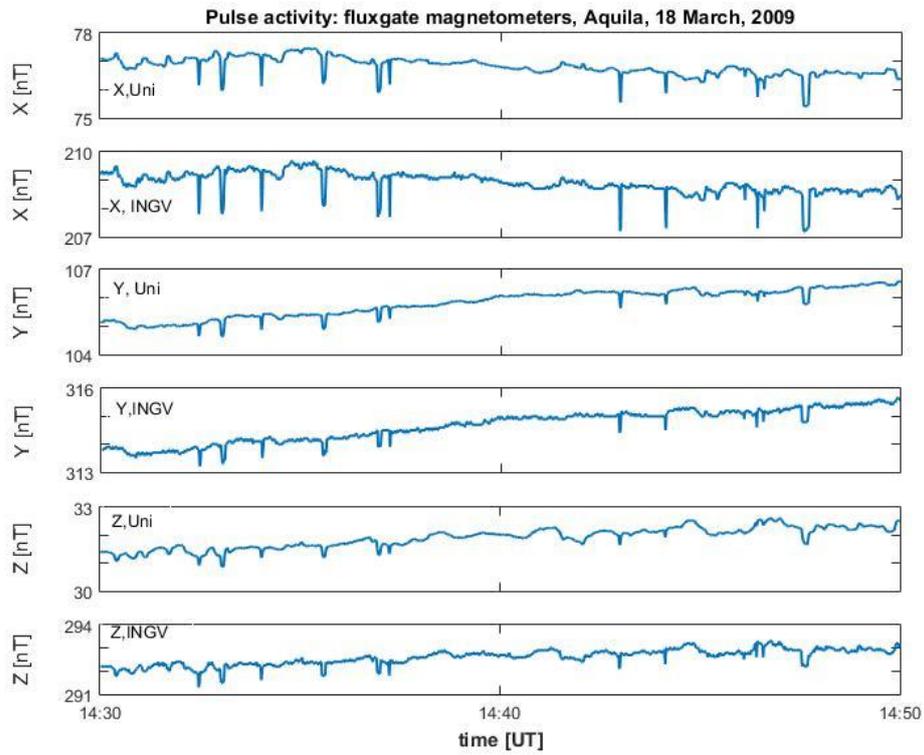

Figure 1.

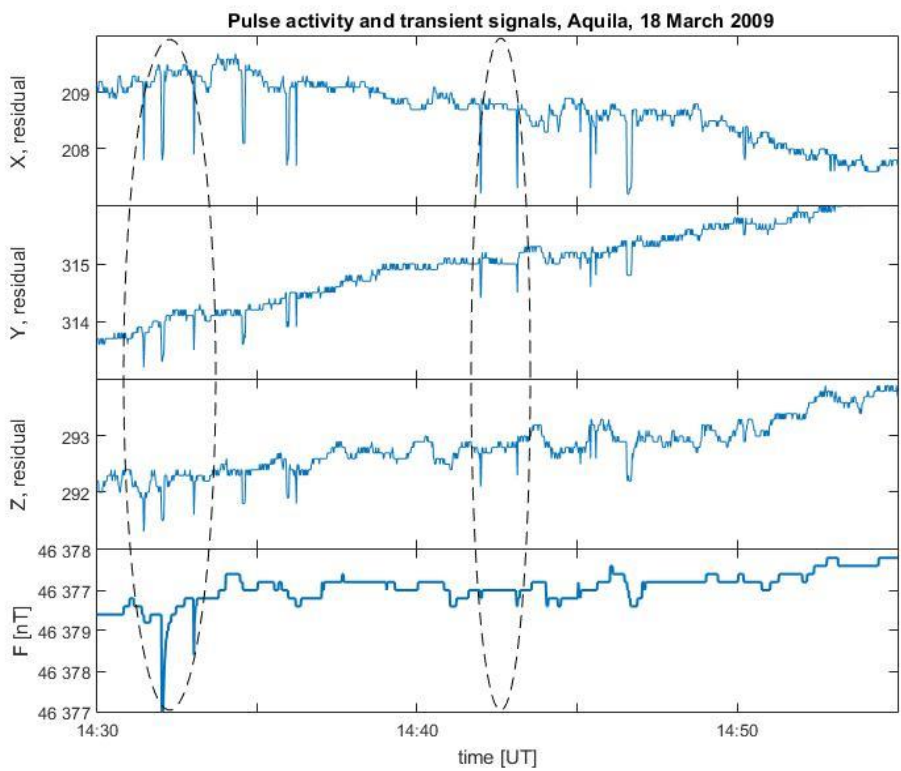

Figure 2.

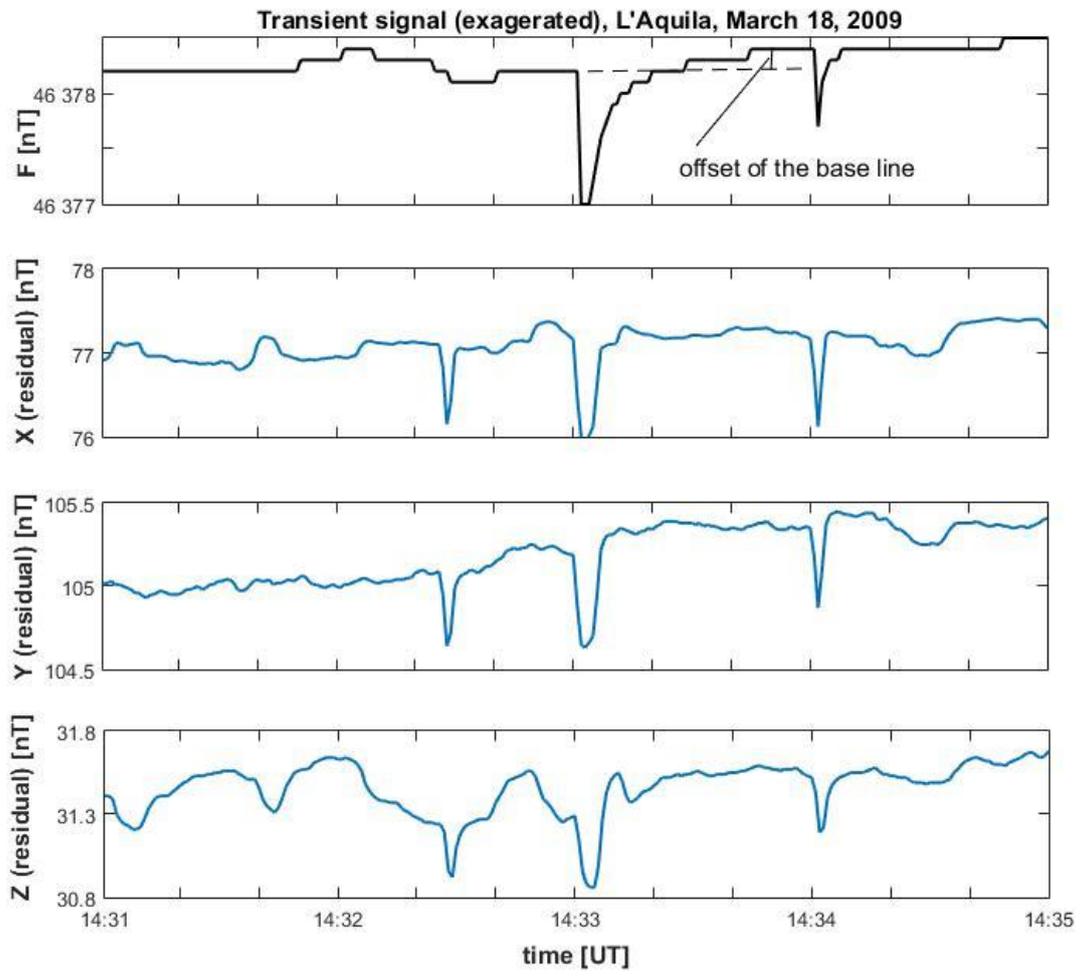

Figure 3.

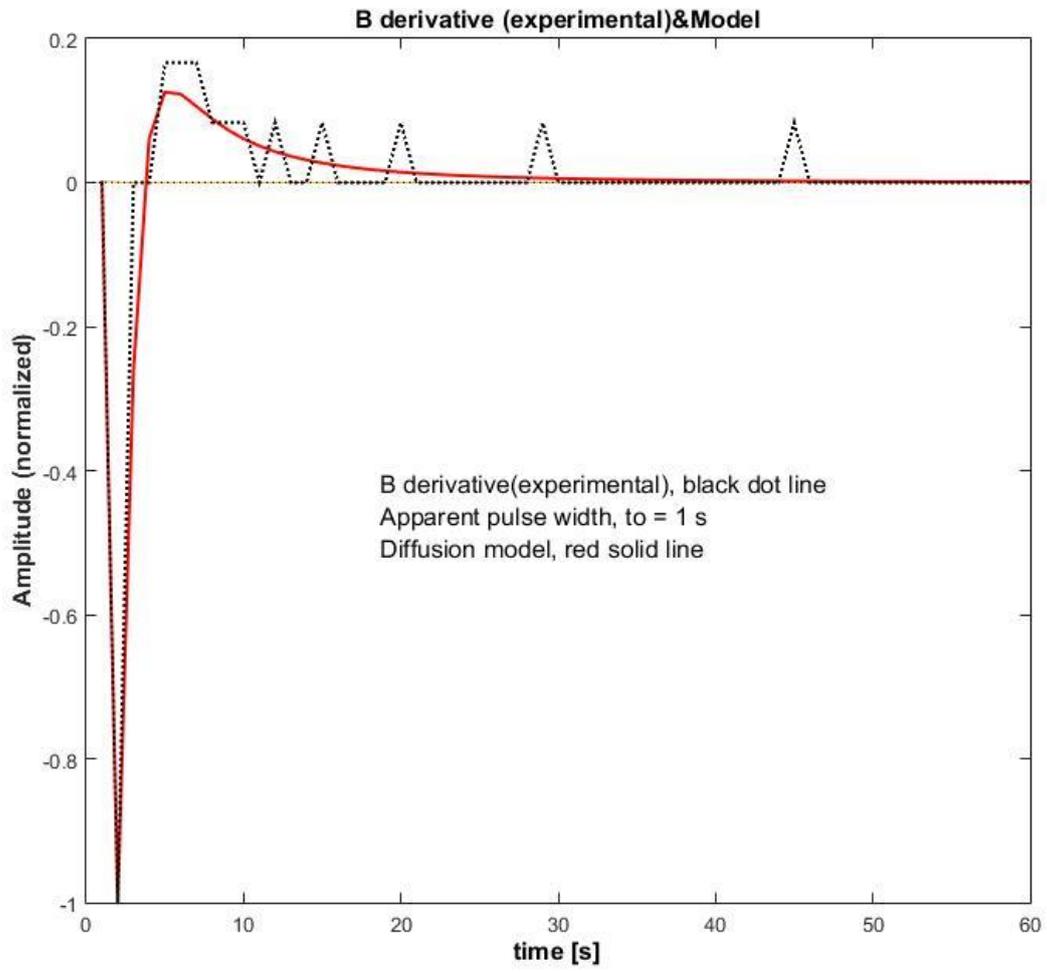

Figure 4.

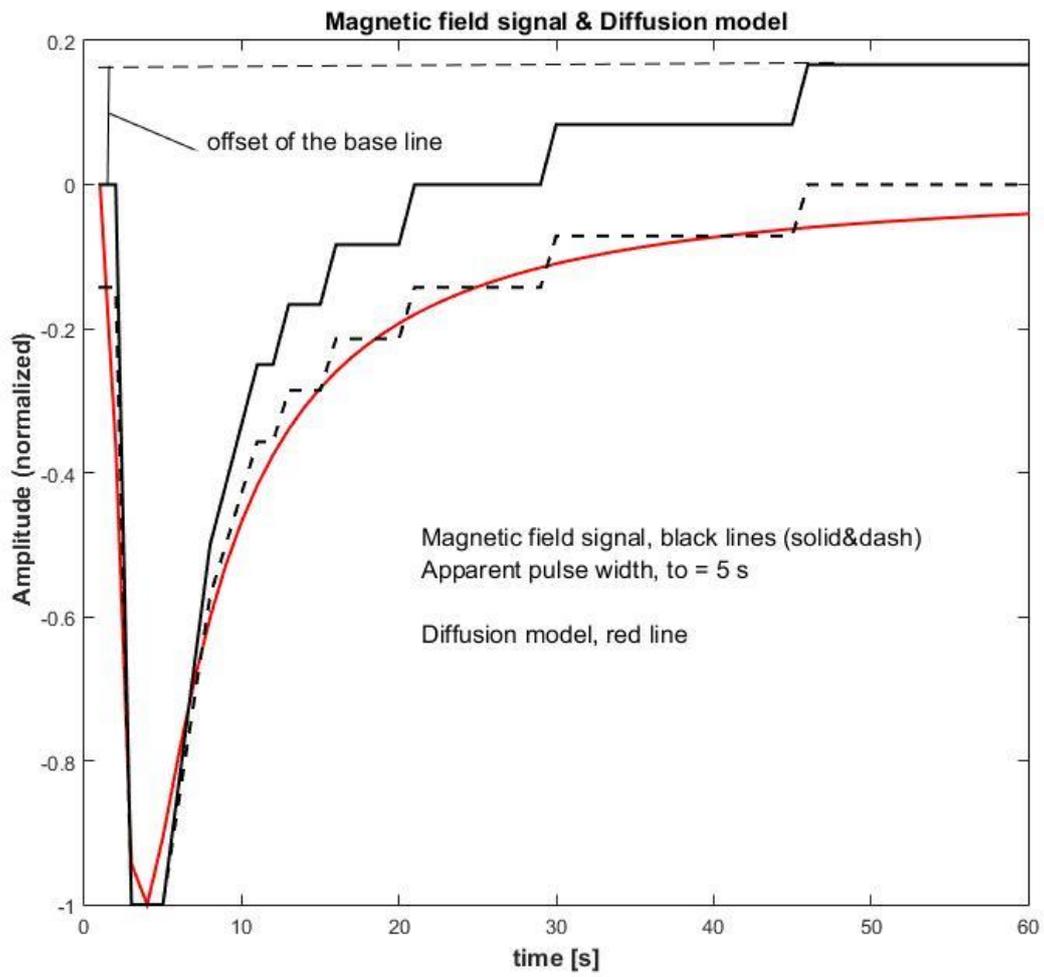

Figure 5.